\documentclass[mypaper,7pt,twoside]{CoAst}
\usepackage{epsf,graphicx,fancyhdr}
\renewcommand{\chaptermark}[1]{\markboth{#1}{}}

\newcommand{\apj}{ApJ}
\newcommand{\aap}{A\&A}  %*************Stimmts?
\newcommand{\mnras}{MNRAS}

\def\kms {km~s$^{-1}$}

\newcommand{\kopf}{\small\itshape Comm. in Asteroseismology \\ Contribution to the Proceedings of the Wroclaw HELAS Workshop, 2008}

\newcommand{\Authors}[1]{\begin{center}\normalsize\bf\sf #1 \end{center}}

\renewcommand{\author}[1]{\begin{center}\normalsize\bf\sf #1 \end{center}}
\newcommand{\Address}[1]{\begin{center}\small\sf #1 \end{center}}

\newcommand{\Session}[1]{{\vspace{3mm}\small \noindent  \hspace*{3mm} Session: } #1 \normalsize}

\newcommand{\Objects}[1]{{\vspace{0mm}\small \noindent  \hspace*{3mm} Individual Objects: } \small #1 \normalsize}

	\newcommand{\threeB}{\small STARS - opacity driving, levitation, opacity data \newline}

\renewenvironment{abstract}{\section*{Abstract}\normalsize\sf}{}
\newcommand{\References}[1]{\begin{flushleft}{\large References\\}\vspace*{2mm}\small #1 \end{flushleft}}

\newcommand{\chapterCoAst}[2]{\chapter[\sf\normalsize #1\\ \footnotesize \hspace*{5mm}by #2 \sf\normalsize][]{#1\\}\rhead[\fancyplain{}{\sf\footnotesize \center{#1}}]{\fancyplain{}{\sffamily\thepage}}\lhead[\fancyplain{\kopf}{\sffamily\thepage}]{\fancyplain{\kopf}{\sf\footnotesize \center{#2}}}}

\renewcommand{\chaptername}{}

\newcommand{\acknowledgments}[1]{\vspace*{5mm}\noindent  \textbf{Acknowledgments.} #1}

\usepackage{journals}

\def\rfr{\smallskip\par\noindent
        \hangindent=7truemm
        \hangafter=1}

\newcommand{\msun}{\mbox{${\rm M}_{\odot}$}}

\begin{document}
\sf

 \chapterCoAst{Input from opacity data in computation of pulsation instability}%paper titel and page heading for even pages
{J.\,Montalb\'an,  and A.\,Miglio} %page heading for odd pages
\Authors{J. \,Montalb\'an,  and A. \,Miglio} 
\Address{Institut d'Astrophysique et G\'eophysique, Universit\'e de Li\`ege,\\ All\'ee du 6 Ao\^ut, 4000 Li\`ege, Belgium}

\noindent
\begin{abstract}
Several types of variable stars are found along the HR diagram whose pulsations are driven by the $\kappa$-mechanism. Given their nature, the precise ($T_{\rm eff}-L$) domain where these pulsators are located is highly dependent on the value of opacity and on its variation inside the star. We analyze the sensitivity of opacity driven pulsators of  spectral-type A and B  ($\delta$~Scuti, $\beta$~ Cephei and SPB stars) to the opacity tables (OP/OPAL) and to  the chemical composition of the stellar matter. We also briefly discuss the effect of opacity on  pulsators whose oscillations are not driven by the $\kappa$-mechanism, such as $\gamma$~Doradus and solar-like stars.
\end{abstract}

\Session{ \threeB }
\Objects{44\,Tau} 

\section*{Introduction}

Already in 1919  Eddington  proposed that an increase of the absorption coefficient in some parts of the star at the moment of maximum compression could excite stellar pulsations. However,  it was necessary to wait around  forty years before  Zhevakin~(1959, and references therein) identified the region of the second ionization of He as responsible of pulsation excitation in  cepheid variables. Baker \& Kippenhahn~(1962) made the first numerical  stability computations applied to realistic models of stellar  envelope  and confirmed that the increase of opacity in the region of second ionization of He leads to sufficient excitation to overcome the damping in the rest of the star, explaining in this way the pulsation of Cepheids.
 Since then, other pulsators have been observed whose excitation mechanism is linked to the variation of opacity with temperature and pressure, not only in the region of second ionization of helium, but also in the Fe-group opacity bump  at $\log T\sim5.3$ ($\beta$~Cepheids and SPB's eg. Dziembowski et al. 1993; sdB's, Kilkenny et al. 1997,  Charpinet et al. 1996) and at $\log T\sim6.25$ (Wolf-Rayet, Townsend \& MacDonald 2006).

The necessary conditions for the excitation of a pulsation mode of period $\tau$ by the $\kappa$-mechanism were described by  Pamyatnykh (1999):
{\it i)} the opacity perturbation in the high temperature phase must grow outward; {\it ii)} the amplitude of the pressure eigenfunction must be large and change slowly with radius  in the driving region; {\it iii)}  the thermal relaxation time ($4 \pi r^3 \rho C_p T/L_R$) in the driving region must be of the order or longer than the oscillation period ($\tau$). The behavior of stellar opacity as a function of temperature and density is not only important in the framework of the first condition.  Stellar opacity determines the density distribution of the equilibrium model and hence the oscillation frequencies. Moreover, the details of the stellar structure are in turn important in the balance between excitation and damping of oscillations. Finally,  opacity of stellar plasma influences the location of convective regions, and hence may play a relevant role in the properties of oscillations.
% of solar-like and  $\gamma$~Dor stars. 

The main ingredients in the computation of stellar opacities are two: atomic physics, and chemical composition of the stellar plasma. In this paper we  examine  the consequences of recent updates on the oscillation properties of different types of main-sequence  pulsators.

\section*{Effect of opacity on stellar structure and  evolution}

\subsection*{Opacity tables}
The improvements included in the new OP (Opacity Project) opacity computations as well as a thorough comparison with OPAL (Iglesias \& Rogers, 1996) opacity tables are described in Badnell et al.~(2005). These changes result in an enhancement of  Rosseland mean opacity, $\kappa_{\rm R}$, at high density and temperatures, and  an increase of 18\% of opacity in the Z-bump due to the new Fe atomic data. The new OP opacity tables are much closer to the OPAL ones  than the previous release (Seaton et al. 1994).  Some differences remain, however, at low temperatures ($\log T < 5.5$): the OP Z-bump in $\kappa_{\rm R}$  presents a hot wing slightly larger than the OPAL one. This translates into an $\kappa_{\rm R}$ difference that can reach 30\% at low densities. On the other hand, OP and OPAL agree at high density within  5-10\%.

\begin{figure*}[ht!]
\centering
\resizebox{\hsize}{!}{\includegraphics{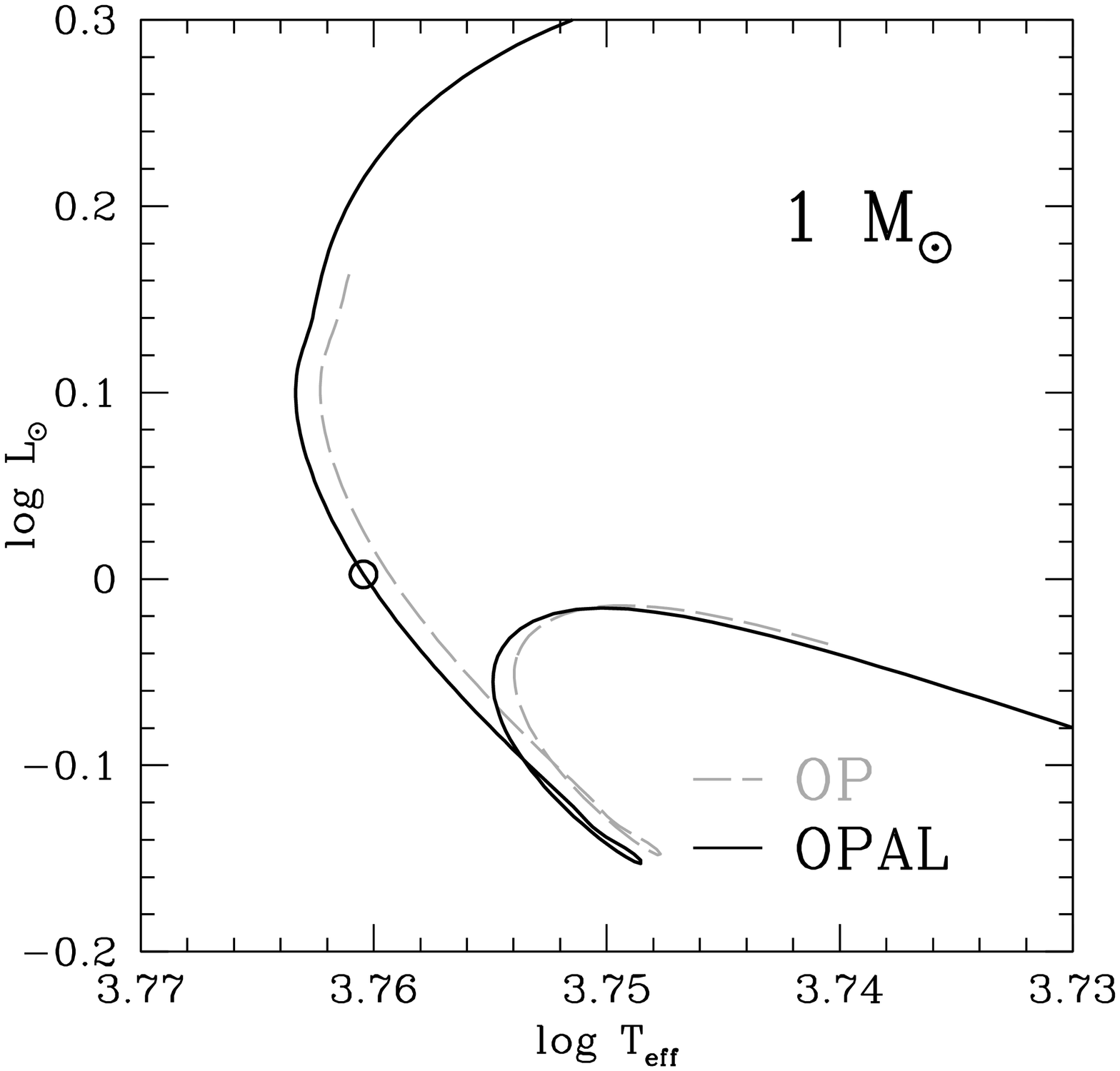}\includegraphics{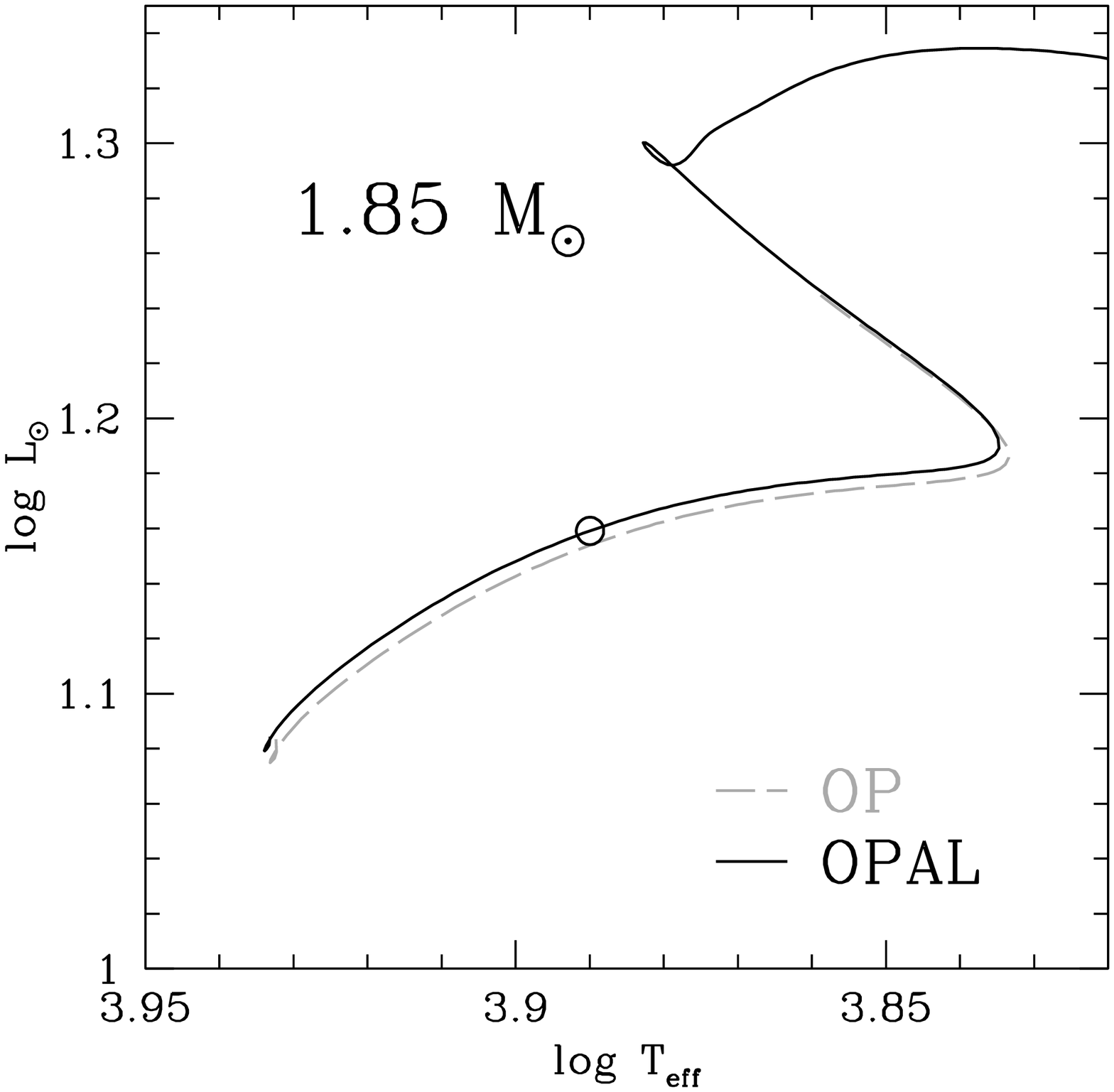}\includegraphics{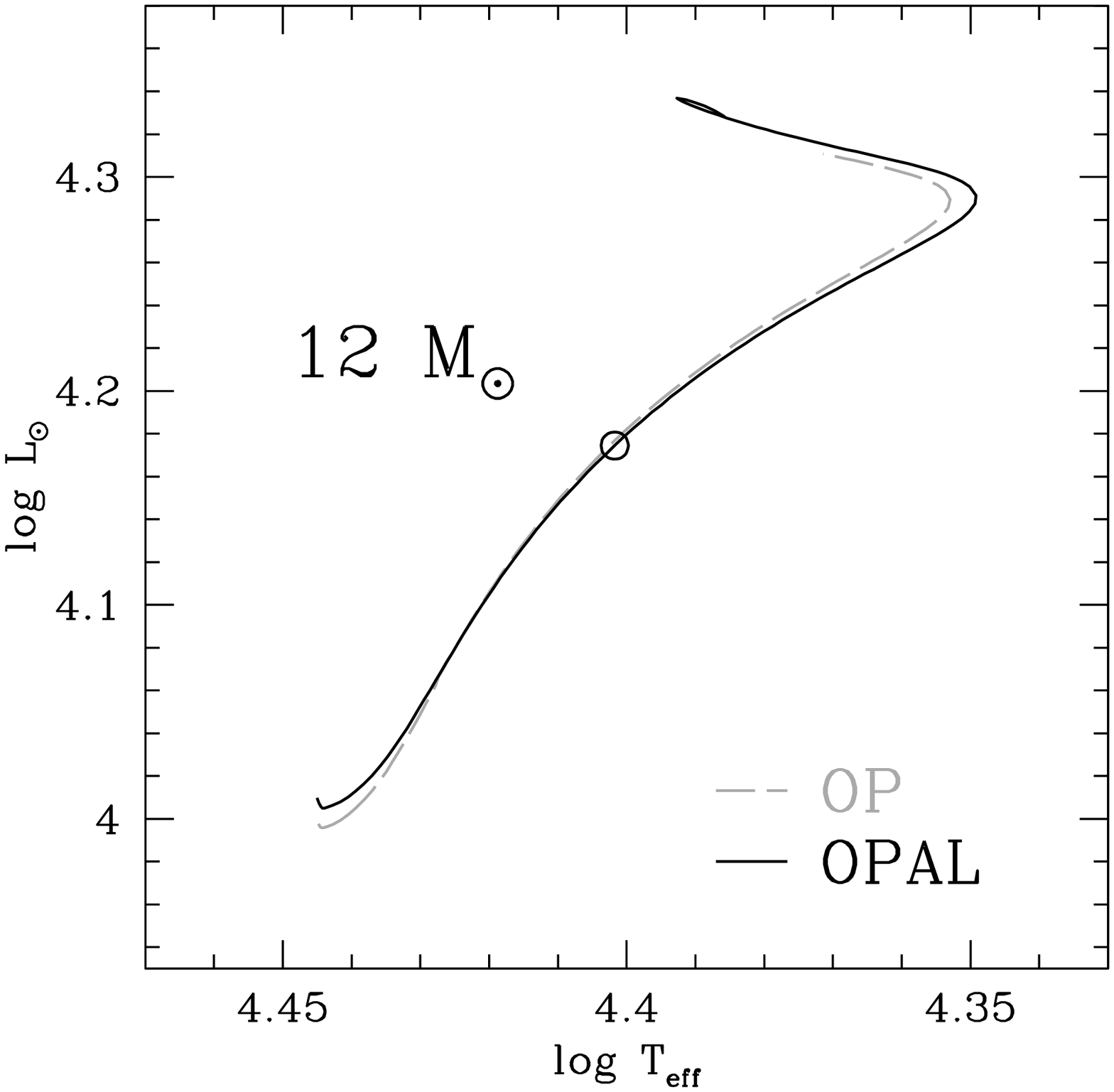}}
\resizebox{\hsize}{!}{\includegraphics{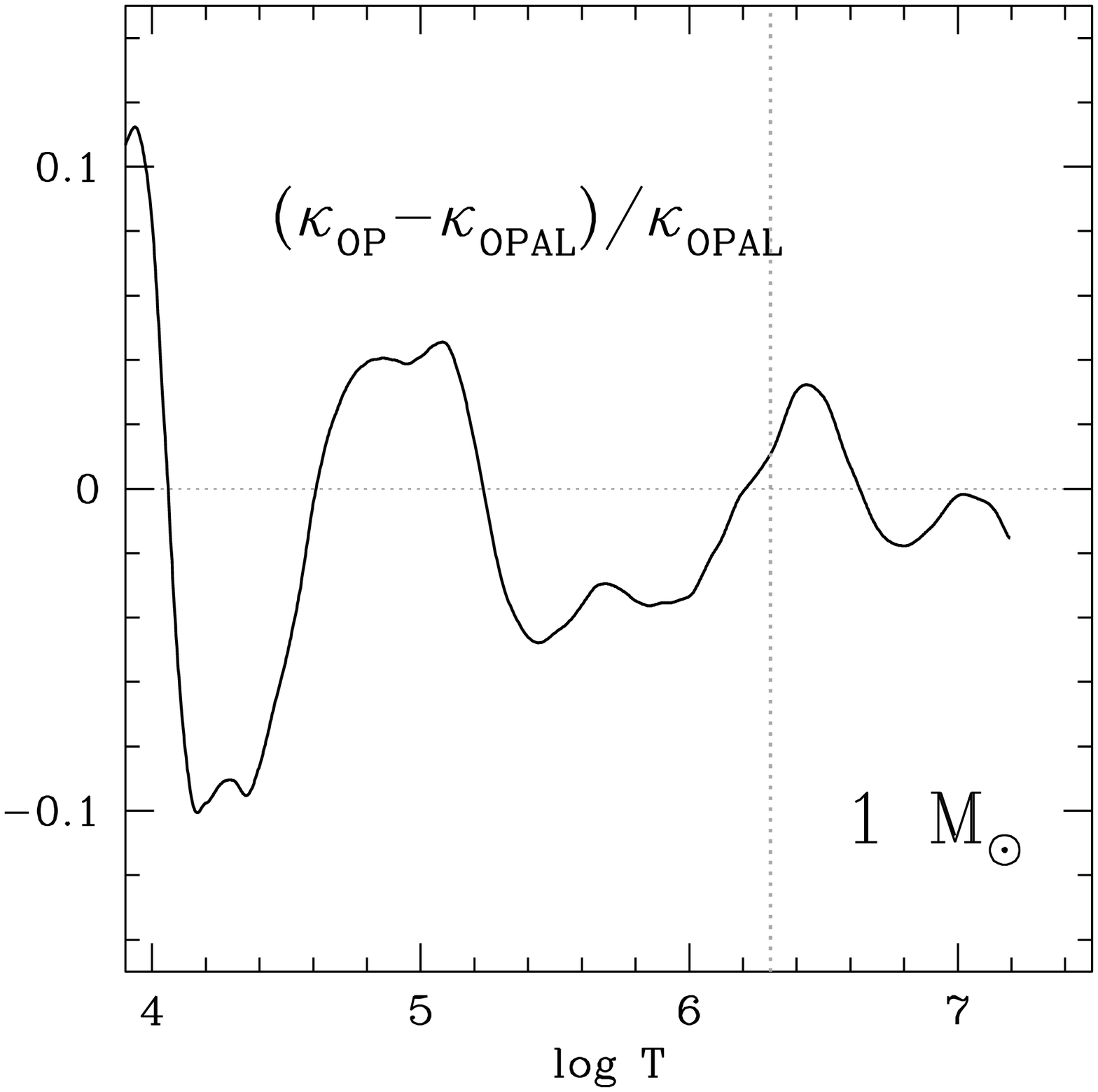}\includegraphics{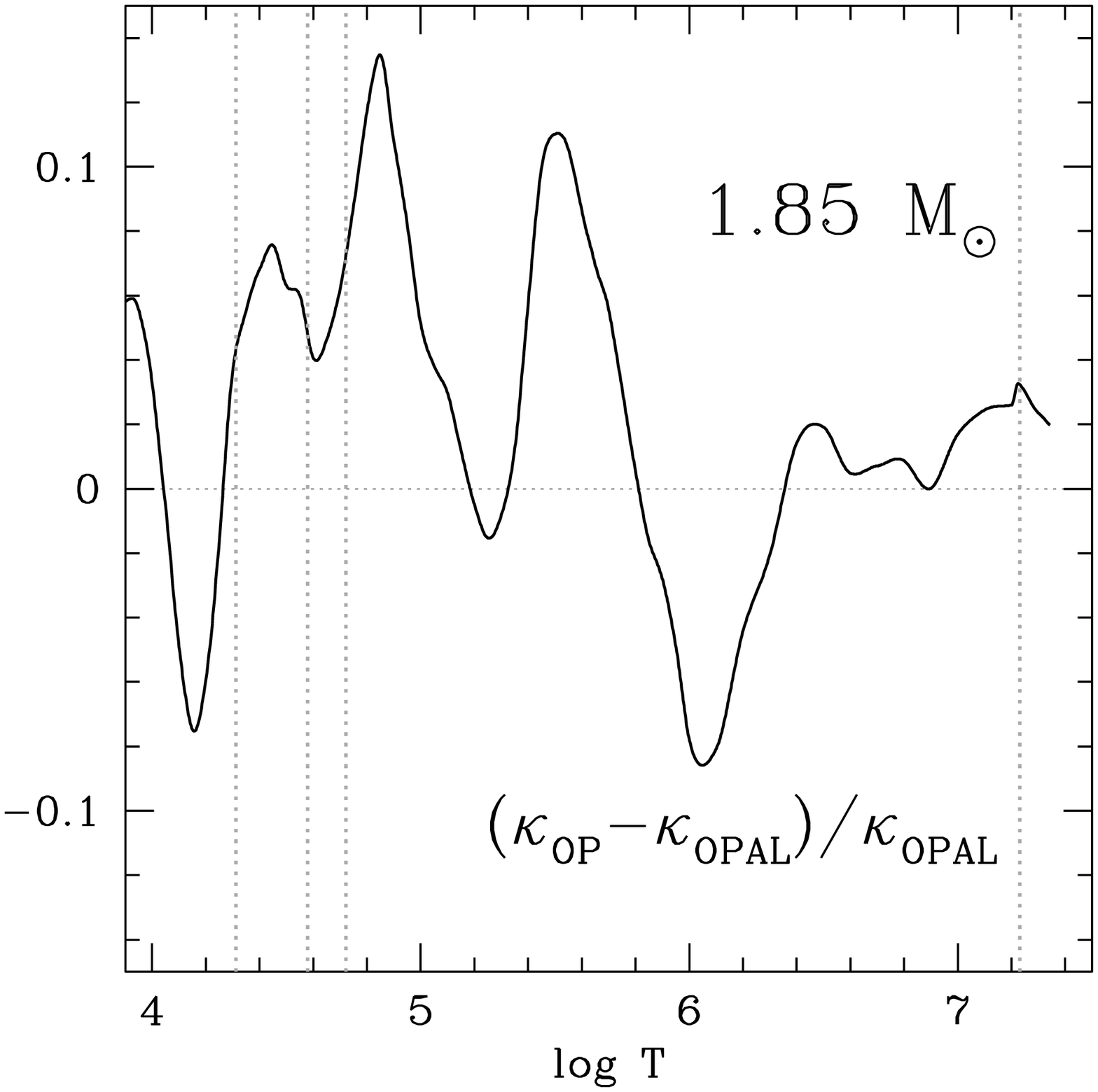}\includegraphics{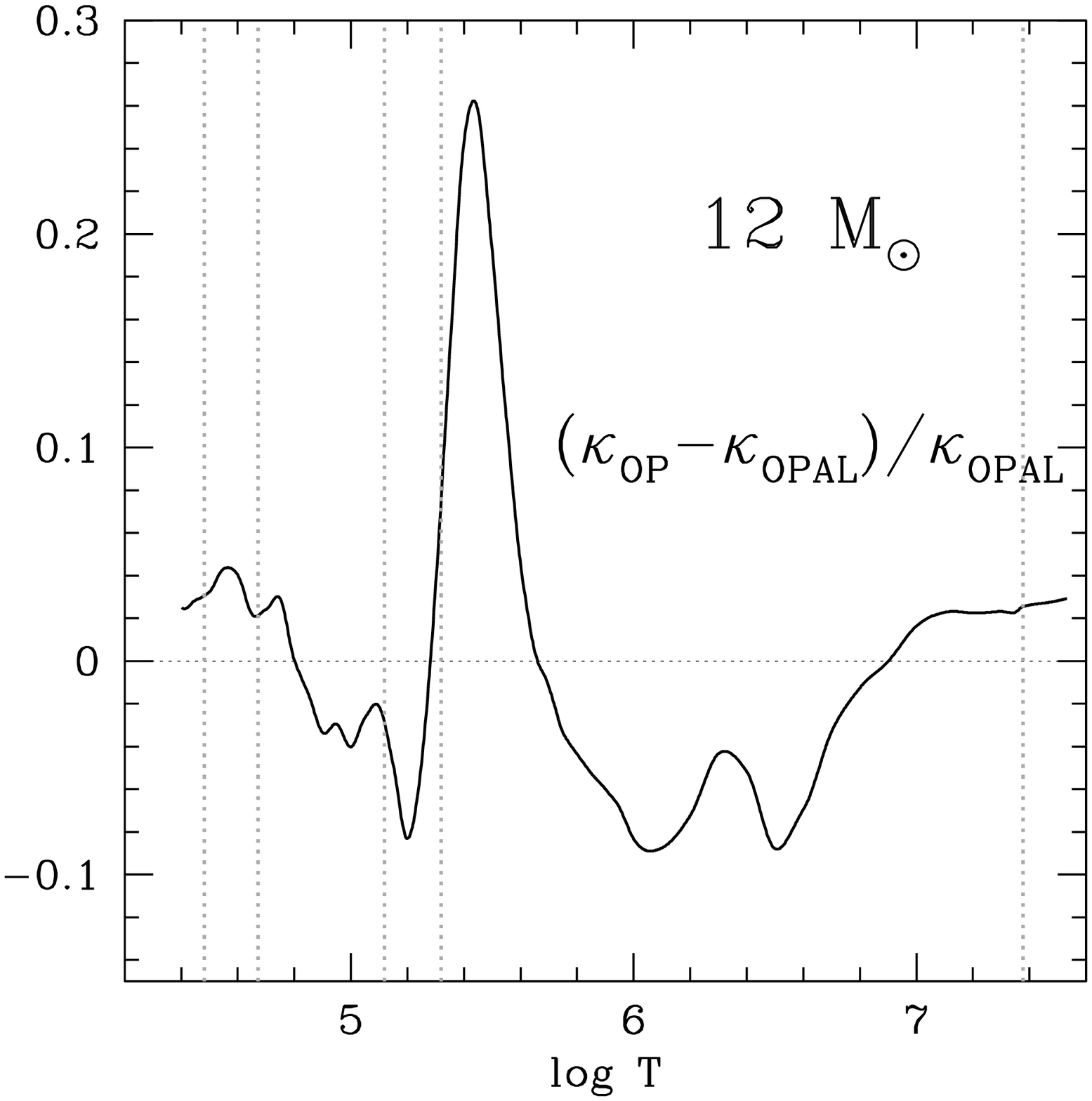}}
\caption{Upper panels: evolutionary tracks of 1~\msun\ (left), 1.85~\msun\ (center) and 12~\msun\ (right) computed with OPAL (solid lines) and OP (dashed lines) opacity tables,  same chemical composition ($X=0.70$, $Z=0.02$) and  treatment of convective transport. Lower panels: differences between OP and OPAL $\kappa_R$ through an  OPAL model at the middle of the MS (mass fraction of hydrogen at the center $X_{\rm c}$=0.35). Vertical lines indicate the boundary of convective regions  in those models.}
\label{fig:OP-OPAL}
\end{figure*}

This disagreement  should affect differently  models of  typical  solar-like, $\delta$-Scuti, and $\beta$~Cephei stars.  In figure~\ref{fig:OP-OPAL} we plot OP and OPAL  evolutionary tracks for models with typical values of mass for these three kind of variables (1.0, 1.85,  and 12~\msun). The  1.0 and 1.85~\msun\  OP evolutionary tracks are cooler than the OPAL ones. Since all the stellar parameters in the models are the same, these results  suggest  a larger OP opacity with respect to the OPAL one. The differences across the  stellar structure of main sequence models are shown in the lower panels of figure~\ref{fig:OP-OPAL}. We emphasize a  difference of  $\sim 30$\% at  $\log T\sim 5.3$ for the 12~\msun\ model, that will  have important consequences on the instability strip of B-type pulsators, and also that OP opacities are systematically 10\% lower than the OPAL ones in the region  around $\log T\sim 6$.

\subsection*{Solar mixture}
\begin{figure*}[ht!]
\centering
\resizebox{\hsize}{!}{\includegraphics{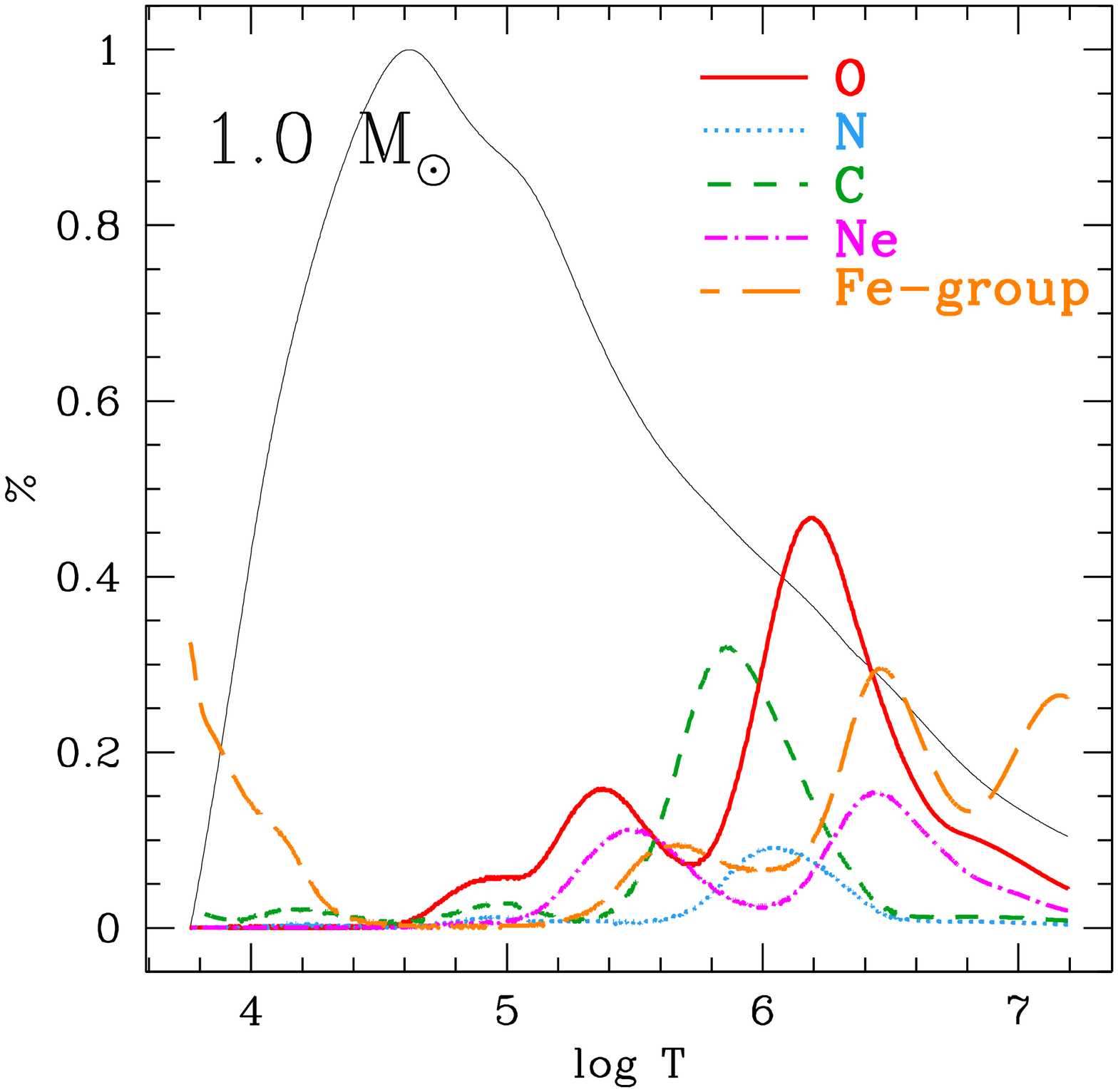}\includegraphics{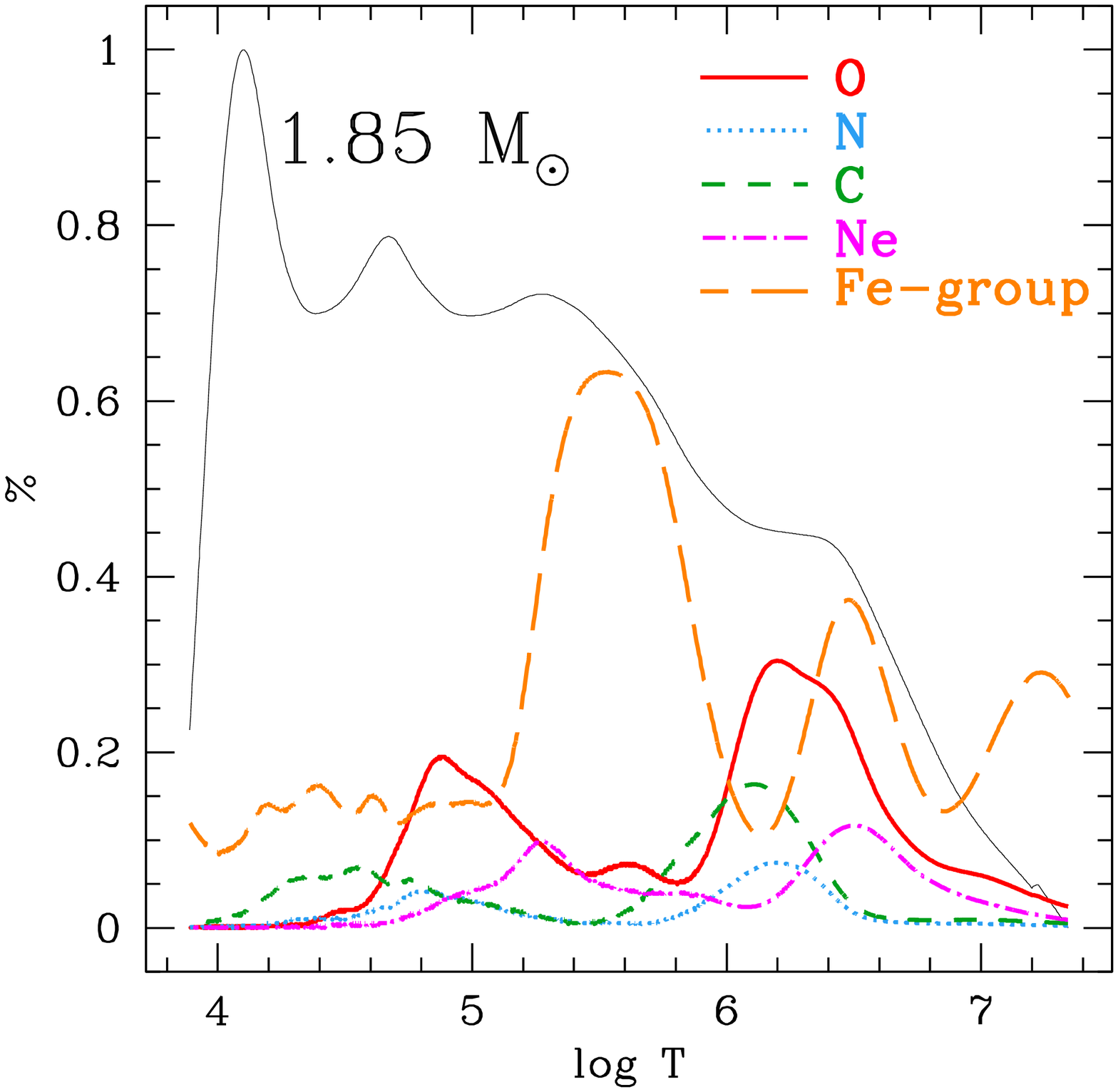}\includegraphics{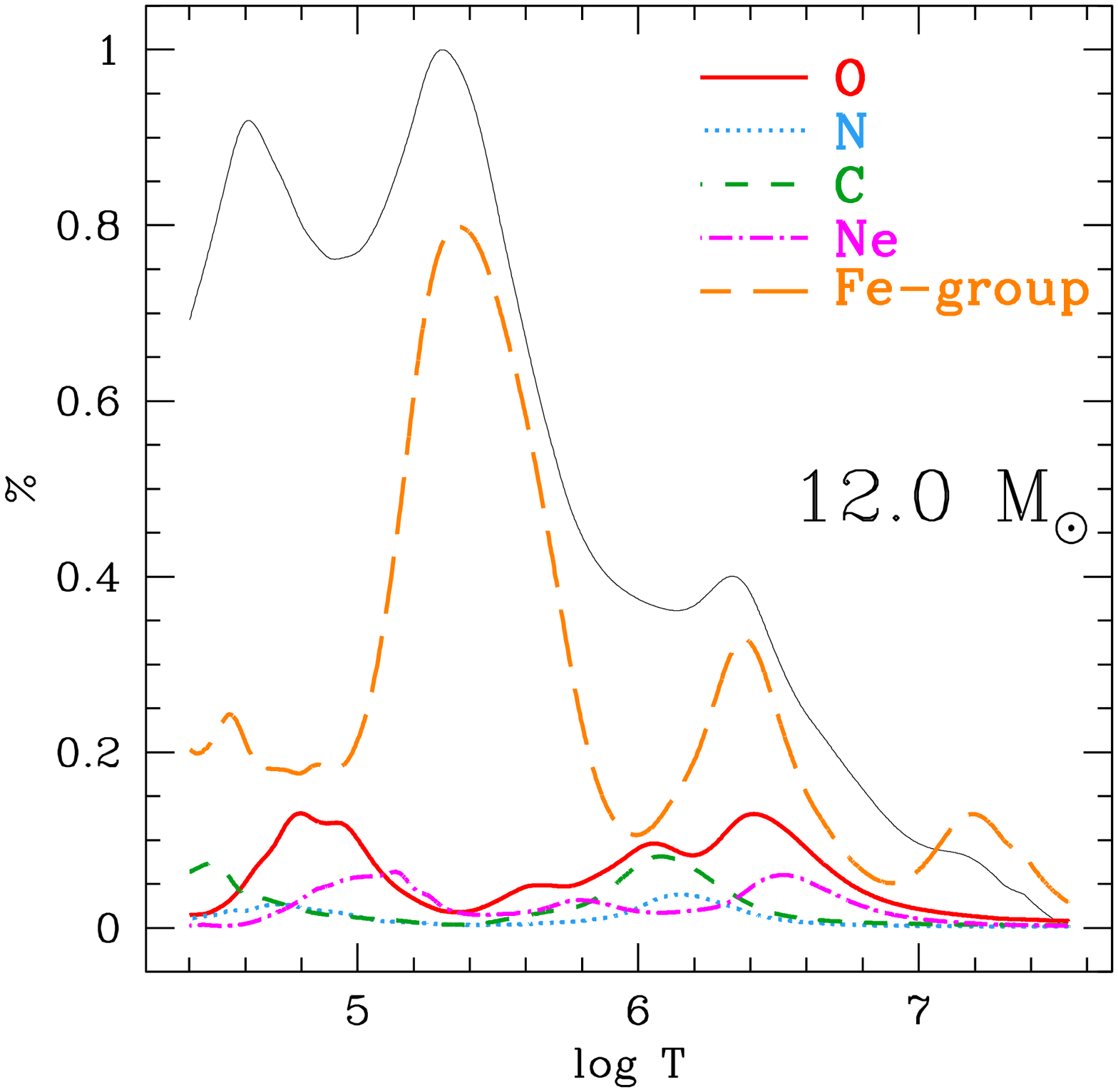}}
\caption{Profile of Rosseland mean opacity (thin solid line) for the same models as in figure~\ref{fig:OP-OPAL}. For each model we derived, by using the OP-sever facility\protect\footnotemark, the contribution to $\kappa_{\rm R}$ in percent coming from O (thick solid lines), N (dotted lines), C (short-dashed lines), Ne (dash-dotted lines) and from  the elements of the iron group (long-dashed lines)}
\label{fig:opac_elem}
\end{figure*}

The re-analysis of solar spectrum by Asplund et al.~(2005) (hereafter AGS05)  including NLTE effect as well as  tri-dimensional model atmosphere computations,  has led to a significant decrease of C, N,  O  and Ne solar abundances and as consequence, to a solar metallicity ($(Z/X)_{\odot}$)  30\% smaller than the value provided by the ``standard''  GN93 mixture (Grevesse \& Noels~1993). The abundance of iron ($\log N({\rm Fe})/N({\rm H})$) has not been affected by the new analysis, which means an iron mass fraction  in AGS05 mixture 25\% larger in than in GN93 one for a given $Z$. It is worth noting that while metallicity is mainly determined by the abundance of C, N and O, the value of the Rosseland mean opacity can be  strongly affected by other less abundant elements. The contribution to opacity from each element in $Z$ depends on density and temperature, hence the effects of solar mixture changes on stellar  pulsation properties  are expected to depend on  spectral type.
Figure~\ref{fig:opac_elem} shows the variation of $\kappa_{\rm R}$ with temperature through the structure of MS stellar models with masses of 1.0, 1.85 and 12.0~\msun, and for each of them, the contribution of C, N, O, Ne and Fe-group elements  (Fe, Ni, Co, Mn) to $\kappa_{\rm R}$. For 1.85~\msun\ (typical $\delta$~Scuti star), $\kappa_{\rm R}$ (thin solid line) clearly shows the peaks due to H, and He ionization ($\log T \sim 4.1$), to He$^+$ ionization ($\log T \sim 4.6$), and to the so called Z-bump of opacity at $\log T \sim 5.3$). The opacity peak due to He$^+$ ionization is also evident in $\kappa_{\rm R}$ curve for 12.0~\msun\ model, nevertheless, for this model the main contribution to opacity comes from Fe-group elements.  We also notice that the contribution of CNO  to $\kappa_{\rm R}$ decreases as stellar mass increases.

\footnotetext{http://opacities.osc.edu/rmos.shtml}

\begin{figure*}[ht!]
\centering
\resizebox{\hsize}{!}{\includegraphics{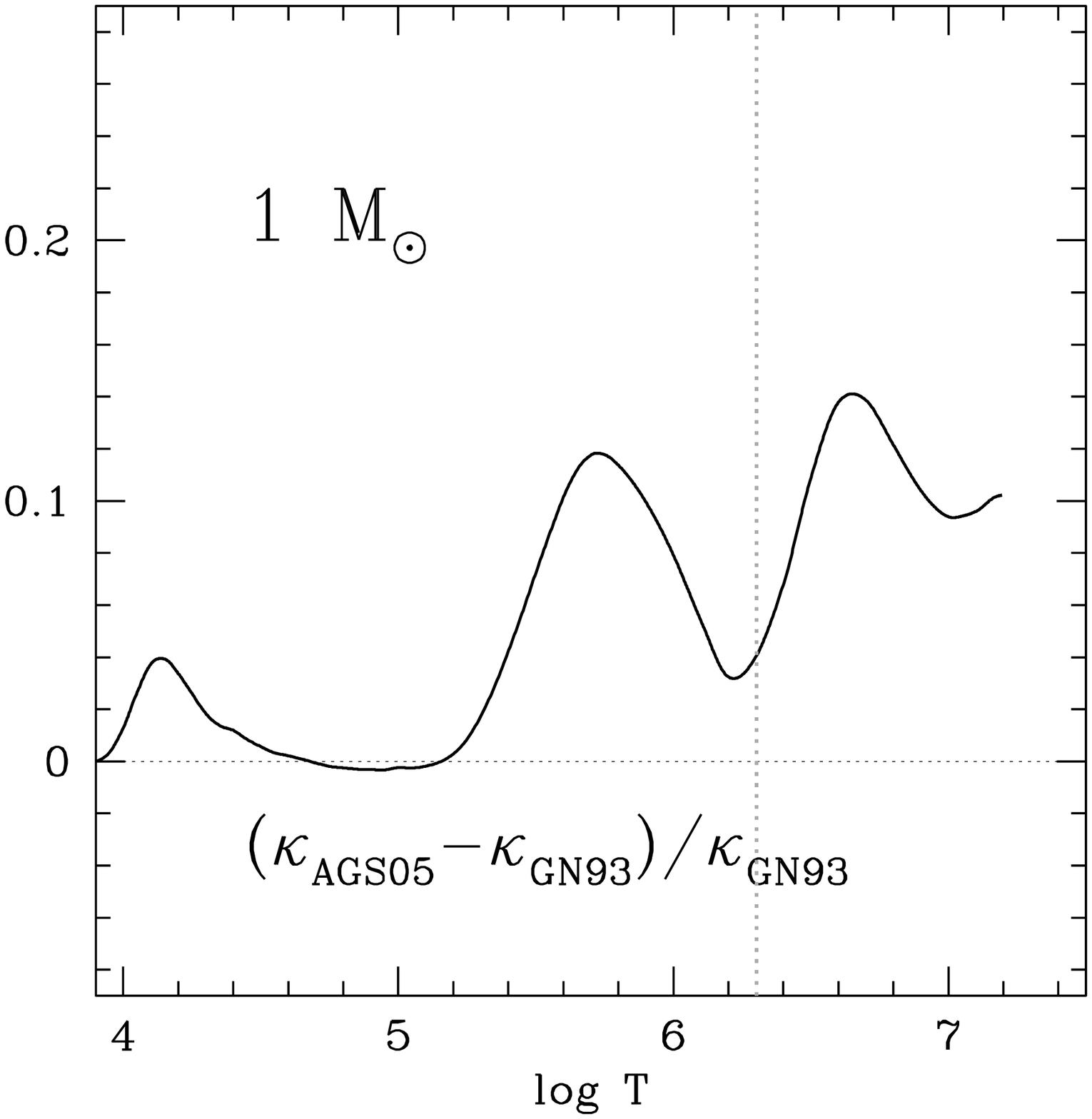}\includegraphics{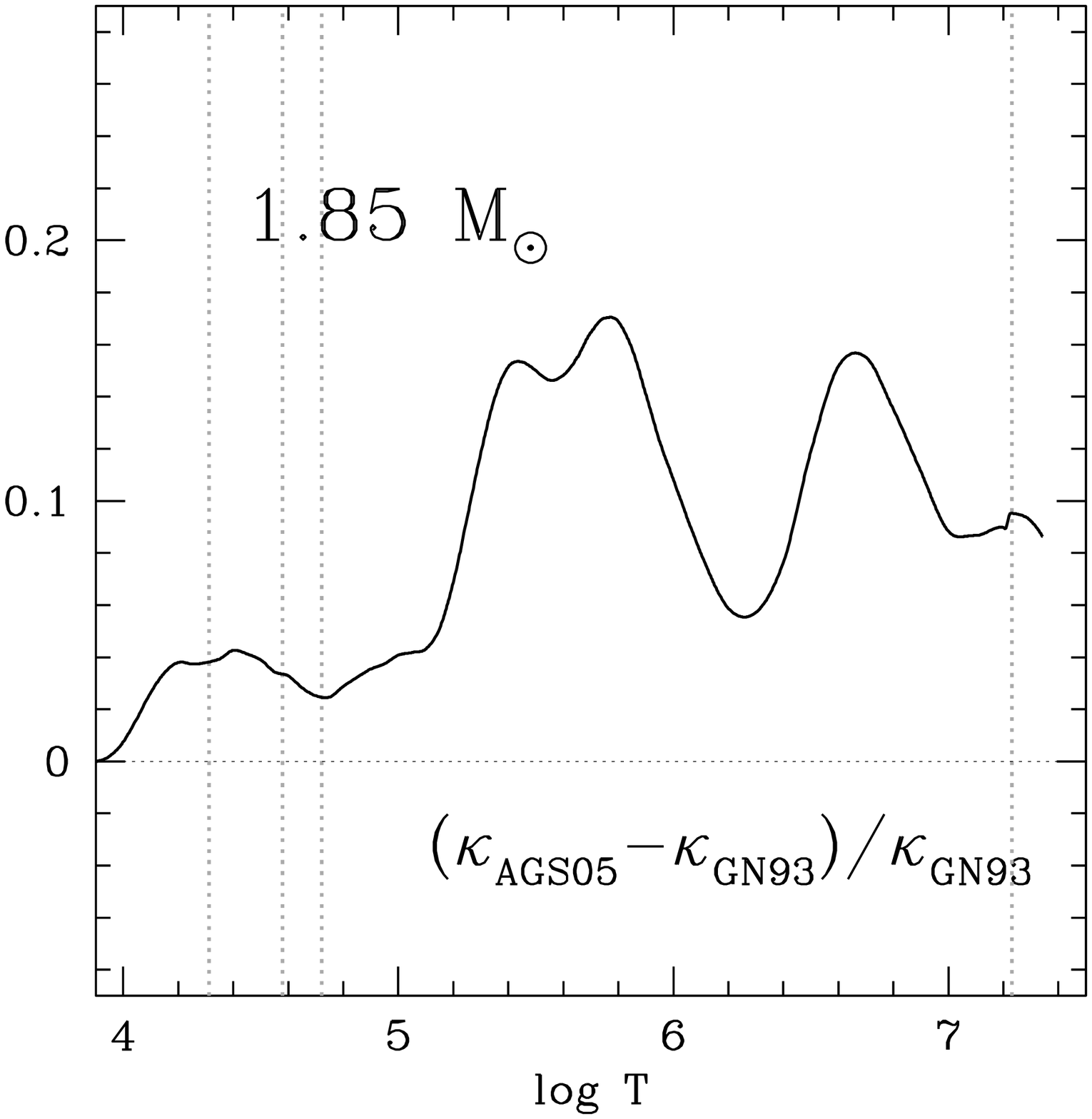}\includegraphics{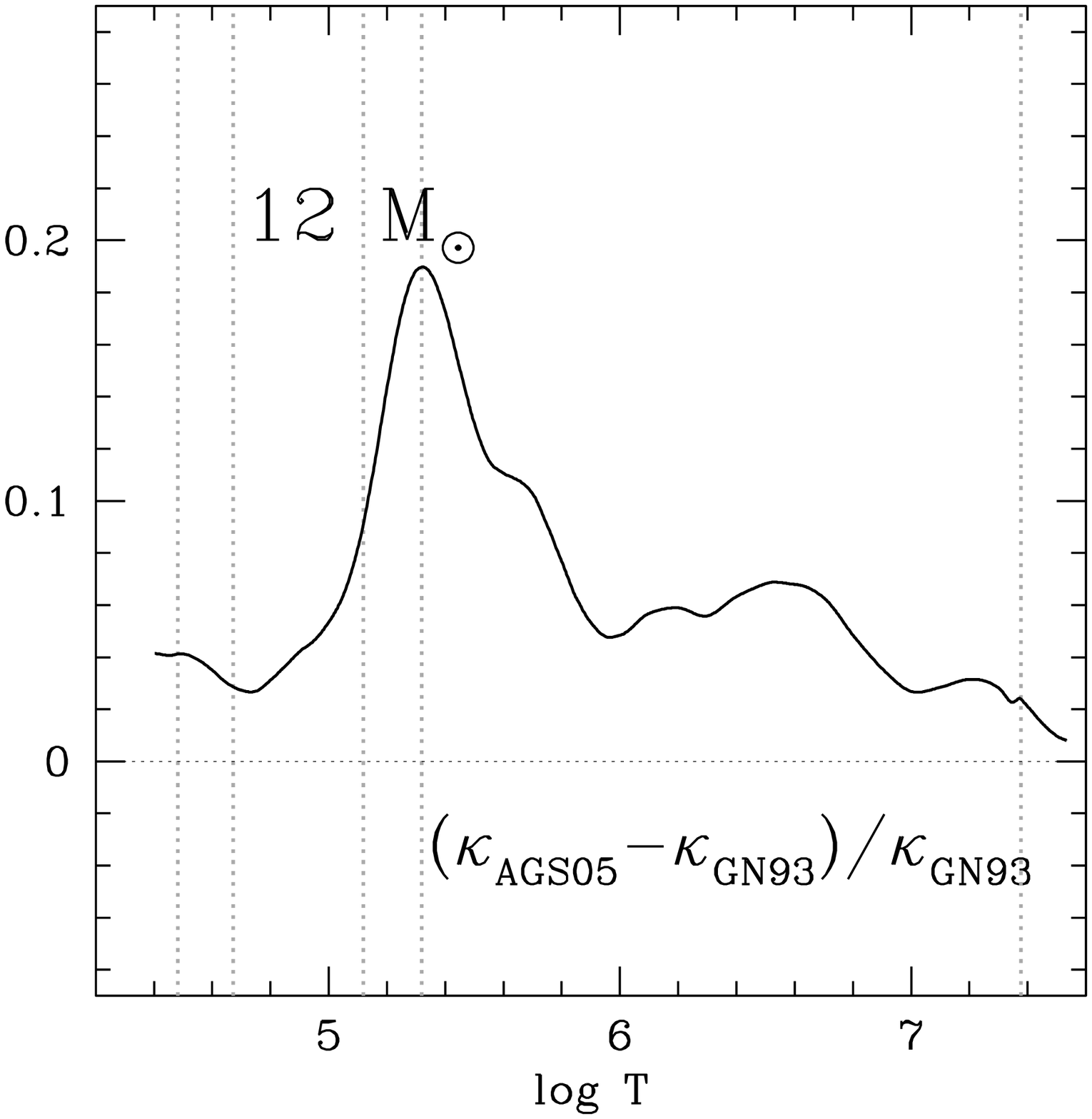}}
\caption{Relative differences of $\kappa_{\rm R}$ values provided by GN93 and AGS05 metal mixtures at fixed $Z=0.02$, through the structure of the GN93 models. The models considered in left, central and right panel are the same as in figures~\ref{fig:OP-OPAL} and \ref{fig:opac_elem}. Vertical lines corresponds to the boundaries of convective regions.}
\label{fig:solarmix}
\end{figure*}

The total effect of new solar mixture on the value of $\kappa_{\rm R}$ is shown in figure~\ref{fig:solarmix}. It should be noticed  that for a given value of metal mass fraction $Z$, AGS05 mixture leads to a larger $\kappa_{\rm R}$ than GN93 one. Nevertheless, it should be kept in mind that the new $Z_{\odot}$ is only 70\% of the old one.   The hollow at $\log T\sim 6.2$ in $\delta \kappa_{\rm R}$ for 1.0 and 1.85~\msun\   (left and central panels  in Fig.\ref{fig:solarmix}) is the signature of the lower oxygen-abundance in AGS05 mixture. While the 30\% decrease of  C,N,O abundances increases the discrepancies between the standard solar model and helioseismology, AGS05 values are in  a better agreement with spectroscopic measurements in B-type stars (Morel et al. 2006).  Moreover, since  Fe-group elements are the main contributors to opacity for these stars,  the  increase of the iron mass fraction by 25\% in the AGS05 mixture will favorably affect the excitation of $\beta$~Cep and SPB pulsation modes in early-type stars.

\section*{Opacity driven  pulsators}
\subsection*{B-type pulsators}

$\beta$~Cep  and Slow Pulsating B (SPB) stars (spectral type B0.5--B3 the former, and B3--B8 the latter)
 are MS pulsators  excited by the $\kappa$-mechanism due to the Fe-group opacity bump at $T\sim 2\times10^5$~K (e.g. Dziembowski et al., 1993). OP opacity values in the Z-bump are larger than OPAL ones by almost 30\% for a typical $\beta$~Cep, moreover  the increase by 25\% of iron mass fraction  for a given $Z$ implies a higher and larger peak in $\kappa_{\rm R}$ at $T\sim 2\times10^5$~K. The combined effect of new opacity and solar mixture on the  SPB and $\beta$~Cephei instability strip has been examined by  Miglio et al.~(2007a,b) and Pamyatnykh (2007).  Their results can be summarized as follows:
while the different profile of $\kappa$ in OP and OPAL computations modifies the blue border of the instability strip, a larger Fe-mass fraction in the metal mixture provides  slightly wider instability bands, and this effect increases as the metallicity decreases. Thus, the number of $\beta$~Cep pulsators expected with AGS05 is more than three times larger than with GN93, and the SPB instability strip resulting from  OP calculations  is 3000~K larger than   with OPAL ones.  
As a consequence, the number of expected hybrid $\beta$~Cep--SPB objects is also larger for OP models.
 The Fe-mass fraction enhancement in the AGS05 mixture, compared with GN93, has the main effect of extending towards higher overtones the range of excited frequencies.
While for $Z=0.01$ the increase of B-type pulsators resulting from the updates of physics in the models is remarkable, for $Z=0.005$  SPB-type modes are excited when considering OP with AGS05, but  none of the different OP/OPAL and GN93/AGS05 evolutionary tracks for masses up to 18~$M_{\odot}$ predicts $\beta$~Cep pulsators.

The updates in basic physics  lead hence  to decrease the disagreement between theory and observations. In fact, the presence of B-type pulsators in low-metallicity environments (Kolaczkowski et al 2006, Karoff et al. 2008, Diago et al. 2008), as well as that of  hybrid  $\beta$~Cep--SPB pulsators  (Handler et al. 2006; Jerzykiewicz et al 2005) cannot be explained in the context of OPAL opacities and GN93 mixture. Nevertheless, some difficulties remain, for instance,  on how to explain the presence of $\beta$~Cep pulsators observed in the Small Magellanic Cloud (Kolaczkowski et al 2006), and the pulsation spectrum detected in $\nu$~Eri (Dziembowski \& Pamyatnykh 2008).

The understanding of pulsations in  subluminous B stars (sdB) has also benefited from  the updated opacity values (Jeffery \& Saio 2006), showing that B-type pulsators can be considered as a ``critical test for stellar opacity''.

\subsection*{$\delta$ Scuti stars: 44~Tau}

$\delta$~Scuti stars (A-F spectral type) are located in the low luminosity end of the classical instability strip. They pulsate in low order p  and mixed g-p modes excited by the $\kappa$-mechanism in the second ionization region of helium, at  $\log T\sim 4.6$. In a recent paper Lenz et al.~(2008) have analyzed the effects of solar mixture and opacity tables in the seismic modeling of 44\,Tau, a slowly rotating  $\delta$~Sct variable that shows radial and non-radial pulsation modes (Antoci et al. 2007; Zima et al. 2007).  The values of periods of fundamental ($\Pi_0$) and first overtone  ($\Pi_1$) radial modes  together with the low rotational velocity ($v\sin i < 5$~\kms) allow Lenz et al. (2008) to use the Petersen diagram ($\Pi_1/\Pi_0\, vs \log \Pi_0$) to derive the fundamental parameters of 44\,Tau and to analyze their dependence on the basic physics.
Comparison between OP and OPAL opacities through a typical $\delta$~Scuti structure (central panel of fig.~\ref{fig:OP-OPAL}) does not show a significant difference (lower than 5\%) in the driving region. Nevertheless, the authors found that  for the case of 44\,Tau the OPAL opacities are preferable to the OP ones, since  it was not possible to fit the radial modes of 44\,Tau  at the same time than its location in the HR diagram by using OP models.

The Petersen diagrams (Petersen \& Jorgensen, 1972) are known to be very sensitive to the metallicity, and hence to  opacity. To clarify the results obtained by Lenz et al.~(2008) we  compared OPAL and OP stellar models with the same mass (M=1.875~\msun, such as derived by Lenz et al. 2008) and at the same evolutionary stage ($X_{\rm c}=0.4$). Their location in the HR diagrams is quite close, their radii differ by 0.05\%, and hence their fundamental period, $\Pi_0$, by 0.16\% ($\log \Pi_0$=-0.8380 instead of -0.8387). The effect on the period ratio is however larger going from  0.7697 for the OPAL model to 0.7722 for the OP one. This difference is an  indication of a lower OP opacity with respect to  OPAL values. In figure~\ref{fig:OP-OPAL} we see that OP is lower than OPAL by a 10\% in the region around $\log T \sim 6.05$.

\begin{figure}
\parbox{5.8cm}{\includegraphics[width=5.5cm,height=5.5cm]{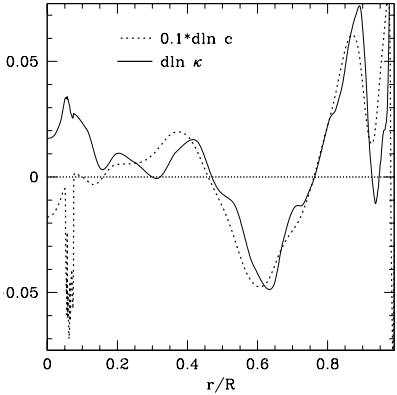}}\parbox{5.5cm}{\caption{ Relative  differences between OPAL and OP $\kappa_{\rm R}$ trough the structure of an OPAL model with $M=1.875$~\msun\, $Z=0.02$, $X=0.70$ and central mass fraction of hydrogen $X_{\rm c}=0.4$. Dotted-line represents the relative difference of sound speed at fixed mass for the two models with the same parameters but different opacity tables (OP et OPAL).}}
\label{fig:eigen}
\end{figure}

 As explained by Petersen \& Jorgensen (1972),  the period ratio ($\Pi_1/\Pi_0$) mainly depends on the effective politropic index ($n=d\ln P/d\ln T-1$) of plasma in a region around $x=r/R=0.7$, much deeper than the second ionization zone of helium. For their stellar envelope models this region corresponds to $\log T$ between 5.5 and 6.2. Figure~4, shows that the largest difference between OP and OPAL opacity values is found  close to $x=0.7$. Different $\kappa_{\rm R}$ involves a different sound speed inside the star and hence different pulsation periods.  A lower opacity  implies larger values of period ratio. Hence, fitting  with OP models the observational $\Pi_0$ and $\Pi_1/\Pi_0$  of 44\,Tau  required  Lenz et al.~(2008) to decrease by 10\% the mass of the model, and therefore the effective temperature and luminosity of the model were not compatible with the  observational error box.

\begin{figure*}[ht!]
\centering
\vspace*{-1cm}
\resizebox{\hsize}{!}{\includegraphics{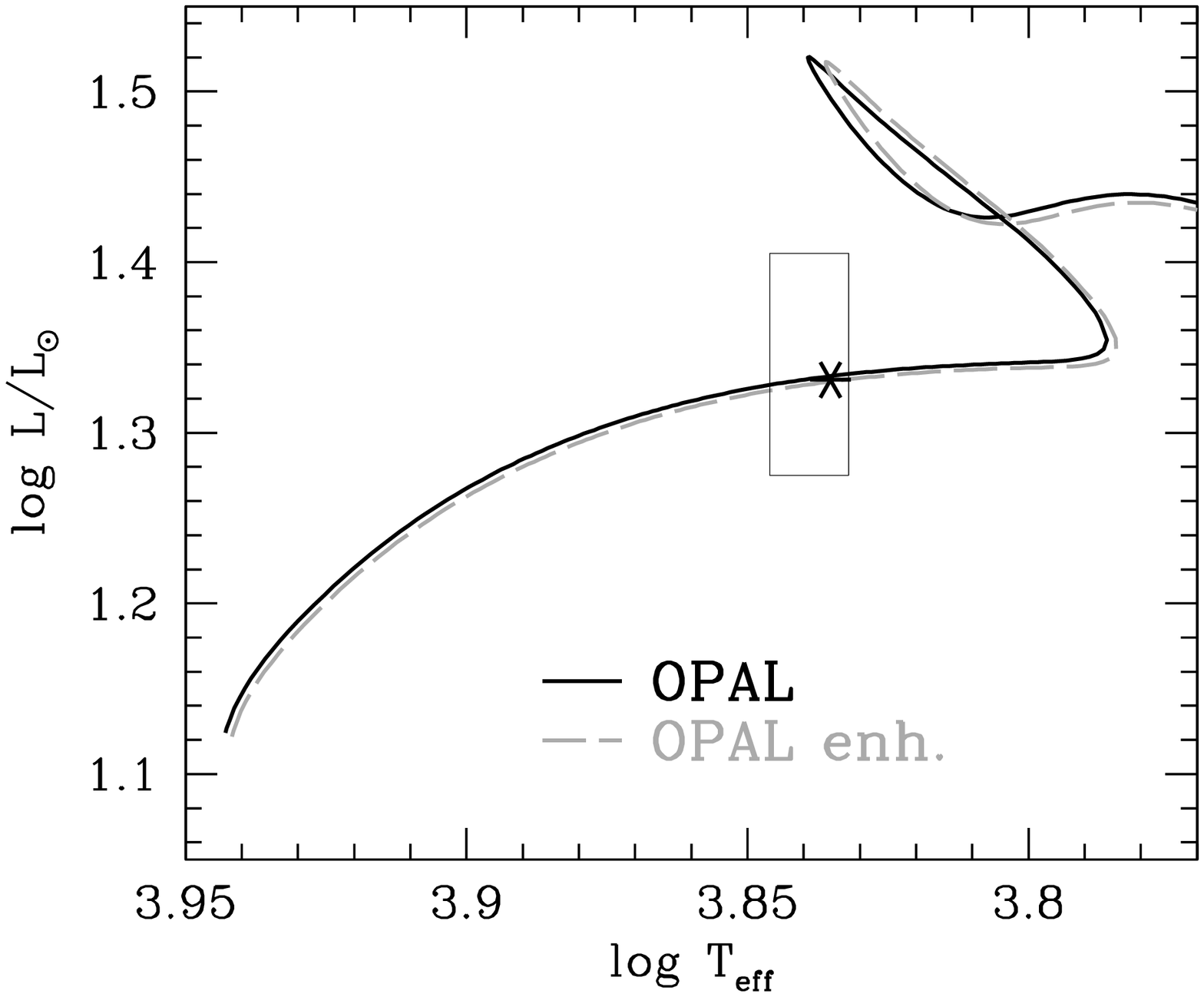}\includegraphics{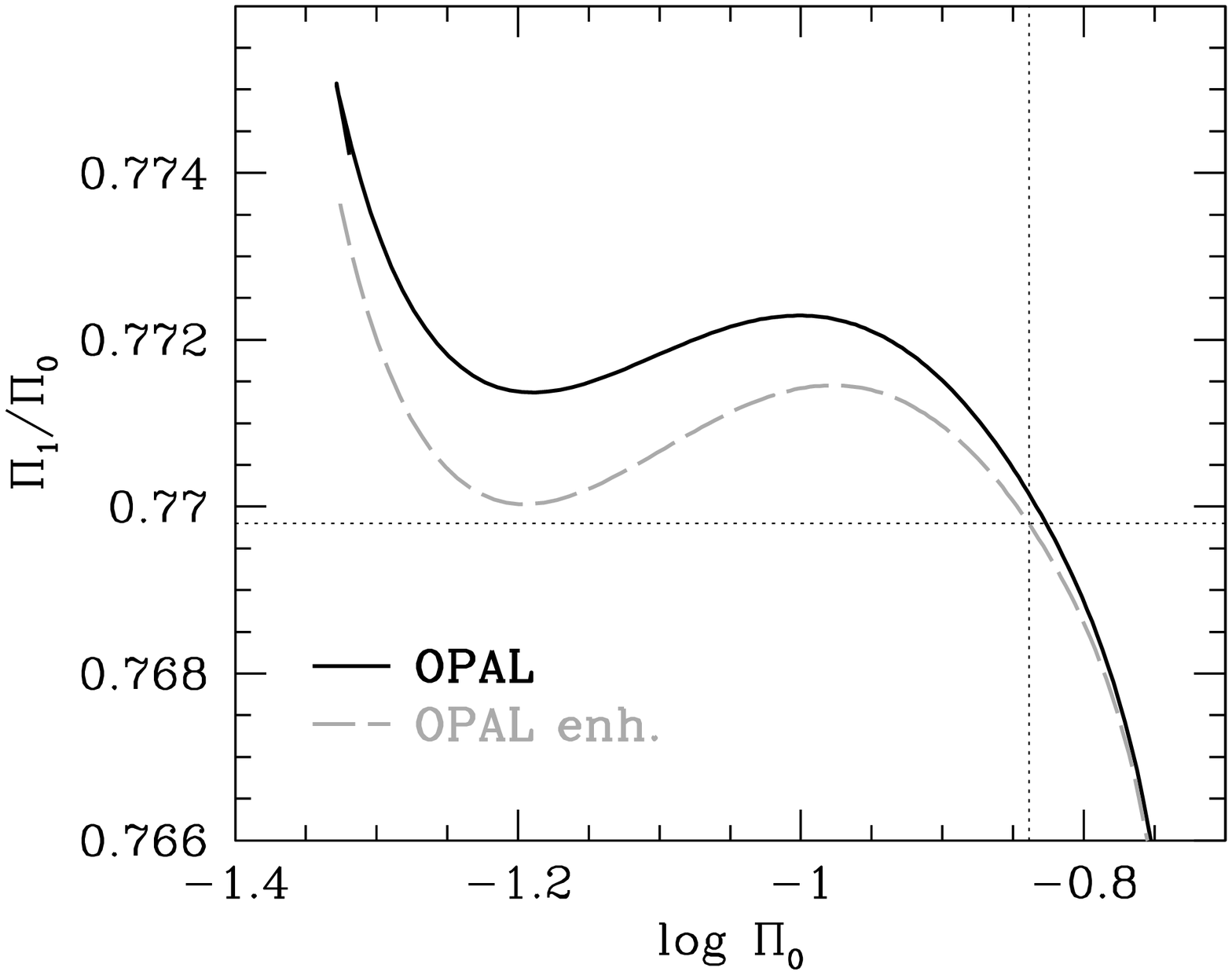}}
\caption{Left panel: evolutionary tracks for $M=1.9$~\msun, $Z=0.02$, $X=0.7$, $\alpha_{\rm OV}=0.3$ computed with OPAL opacity tables (solid line) and with OPAL enhanced by 5\% in the region around $\log T\sim 6.05$ (dashed line). The square corresponds to the observational error box (Lenz et al. 2008), and the star represent the model fitting $\Pi_0$ and the period ratio ($\Pi_1/\Pi_0$). Right panel: Petersen diagram corresponding to the evolutionary tracks in left panel.}
\label{fig:44tau}
\end{figure*}

Given the sensitivity of  the period ratio to the opacity at $\log T \sim 6.05$ we wonder how the stellar parameters derived by using the Petersen's diagram depende on the uncertainty affecting the opacity values. We found that  by increasing the OPAL opacities by at maximum a 5\%  around $\log T\sim 6.25$, a MS model with $M=1.9$~\msun\, $X=0.70$,  $Z=0.02$, $X_{\rm c}=0.195$ and an overshooting parameter $\alpha=0.3$  satisfies  the observational constrains (HR and Petersen diagrams locations)  as shown in figure~\ref{fig:44tau}. This differs from the result found in Lenz et al.~(2008), where only post-MS models were able to fit observations.

\section*{Other pulsators}

\subsection*{$\gamma$ Doradus pulsators}

$\gamma$ Doradus variables are F-type stars oscillating with high-order g-modes (Kaye et al. 1999), and the excitation mechanism has been identified by Guzik et al.~(2000) as being due to the convective blocking of radiative flux. The location of the convective zone (and hence potentially stellar opacity) plays then a fundamental role in these pulsators.  Computations considering the interaction pulsation-convection by Dupret et al. (2005) confirm the first results by Guzik obtained in the frozen convection approximation and shows that there is also an small contribution of the $\kappa$-mechanism coming from the Z-bump of opacity at $\log T \sim 5.3$, close to the bottom of the convective envelope in $\gamma$~Dor stars.
Despite all that, changing  opacity tables and the solar mixture do not significantly affect neither the instability strip nor the oscillation spectrum (Miglio et al. in preparation). Nevertheless, these objects are still not well known, and some surprises could rise in the future.

\subsection*{The Sun}

Oxygen is the main contributor to $\kappa_{\rm R}$ at the bottom of the solar convective envelope ($\log T\sim 6.3$), followed by iron and neon. The decrease by 30\% of solar metallicity resulting from Asplund et al. (2005) analysis has hence led to drop by 20\% the opacity at the bottom of convective zone and to ruin the good agreement  between the standard solar model and helioseismology (see Basu \& Antia 2008 and references therein).
The updated OP opacities imply only a raise of opacity at the bottom of the convective zone at maximum of 5\%, too small to solve the problem of the helioseismic model. A large variety of suggestions have been proposed to recover the lost opacity and hence the agreement with helioseismic results, however none up to now (Basu \& Antia 2008 for review) has provided satisfactory results, and the problem of the solar model remains.
We would like, nevertheless, to recall that the good agreement between the Solar Standard Model (Christensen-Dalsgaard et al. 1996)  and the seismic one was  also damaged (even if the effect was small compared with the actual solar problem) when the updated version of OPAL opacity tables (Iglesias \& Rogers 1996) led to a decrease of opacity of 5\% at the bottom of solar convective region  $\log T \sim 6.3$. The so called S-model was in fact computed with the first delivery of OPAL opacity tables (Rogers \& Iglesias 1992).

\section*{Conclusions}
In the nineties, the OPAL opacity tables signified a revolution in stellar physics. Since then, the progress in observational techniques and numerical computations have highlighted new discrepancies between theory and observations. In spite of recent improvements some problems persist, for instance:  {\it i)} the presence of $\beta$~Cep pulsators in the SMC; {\it ii)} the excitation of high and low frequencies in B-type pulsators such as $\nu$~Eri (Dziembowski \& Pamyatnykh 2008); {\it iii)} discrepancy between solar model and helioseismology; {\it iv)} the remaining 15\% disagreement  between evolution and pulsation masses of cepheids variables (Keller 2008 and references therein). Some of these discrepancies could be reduced by increasing (yes, again...) $\kappa_{\rm R}$ in limited domains of temperature, in particular at $\log T > 6$ (see also Zdrawkov \& Pamyatnykh 2008, these proceedings).

In spite of recent  updates, some differences between OP and OPAL remains, mainly due to the different equation of state used in those approaches. The differences may seem small ($\sim 10$~\%), but we showed here that an enhancement of 5\%  at $\log T \sim 6$ may be able to modify the results of a seismic analysis. Both opacity compilations, OP and OPAL, were computed assuming reasonable approximations for the then  current accuracy of observations. So, perhaps there is some room to increase stellar opacity. In fact,  not all the elements are considered with the same degree of accuracy, and  heavy and low abundance elements have not been taken in consideration since their contributions to $\kappa_{\rm R}$ was estimated to be small (Iglesias et al. 1995) for the solar model, and the computation very time consuming. 

Stellar pulsation theory  has come out to be a powerful tool to probe different aspects of the stellar physics. Today it suggest that perhaps it is time to come back to review the basic input physics in stellar modeling, in particular stellar opacity computations.

%Here comes a figure...
%\figureDSSN{filename}{caption}{label}{!ht}{clip,angle=X,width=XXmm}

%More text...

\acknowledgments{JM. and A.M. acknowledge financial support
from the European Helio- and Asteroseismology Network HELAS,
from the Prodex-ESA Contract Prodex 8 COROT (C90199) and
from FNRS.
%thank you
}

\References{
\rfr Antoci, V., Breger, M., Rodler, F., et al.,.\ 2007, \aap, 463, 225 
\rfr Asplund, M., Grevesse, N., \& Sauval, A.~J.\ 2005, ASP Conference Series, 336, 25
\rfr Badnell, N.~R., Bautista, M.~A., Butler, K.,et al.\ 2005, \mnras, 360, 458 
\rfr Baker, N., \& Kippenhahn, R.\ 1962, Z.A., 54, 114 
\rfr Basu, S., \& Antia, H.~M.\ 2008, \physrep, 457, 217 
\rfr Charpinet, S., Fontaine, G., Brassard, P., \& Dorman, B.\ 1996, \apjl, 471, L103 
\rfr Christensen-Dalsgaard, J., et al.\ 1996, Science, 272, 1286 
\rfr Diago, P.~D., Guti{\'e}rrez-Soto, J., Fabregat, J., \& Martayan, C.\ 2008, \aap, 480, 179 
\rfr Dupret, M.-A., Grigahc{\`e}ne, A., Garrido, R., et al.\ 2005, \aap, 435, 927 
\rfr Dziembowski, W.~A., \& Pamyatnykh, A.~A.\ 2008, \mnras, 385, 2061 
\rfr Dziembowski, W.~A., Moskalik, P., \& Pamyatnykh, A.~A.\ 1993, \mnras, 265, 588 
\rfr Grevesse N., Noels A., 1993, in Hauck B. Paltani S. R. D., ed., La formation des \'el\'ements chimiques, AVCP La composition chimique du soleil
\rfr Guzik, J.~A., Kaye, A.~B., Bradley, P.~A., et al. \ 2000, \apjl, 542, L57 
\rfr Handler, G.,  Jerzykiewicz, M., Rodríguez, E.,et al.\ 2006, \mnras, 365, 327 
\rfr Iglesias, C.~A., Wilson, B.~G., Rogers, F.~J., et al.\ 1995, \apj, 445, 855 
\rfr Iglesias, C.~A., \& Rogers, F.~J.\ 1996, \apj, 464, 943 
\rfr Karoff, C., Arentoft, T., Glowienka, L., et al.,\ 2008, \mnras, 386, 1085 
\rfr Keller, S.~C.\ 2008, \apj, 677, 483 
\rfr Jeffery, C.~S., \& Saio, H.\ 2006, \mnras, 372, L48
\rfr Jerzykiewicz, M., Handler, G., Shobbrook, R.~R., et al.\ 2005, \mnras, 360, 619 
\rfr Kaye, A.~B., Handler, G., Krisciunas, K., et al.,\ 1999, \pasp, 111, 840 
\rfr Kilkenny, D., Koen, C., O'Donoghue, D., \& Stobie, R.~S.\ 1997, \mnras, 285, 640 
\rfr Ko{\l}aczkowski, Z., Pigulski, A., Soszy{\'n}ski, I.et al.\ 2006, MemSAIt, 77, 336
\rfr Lenz, P., Pamyatnykh, A.~A., Breger, M., \& Antoci, V.\ 2008, \aap, 478, 855
\rfr Miglio, A., Montalb\'an, J., \& Dupret, M.-A., 2007a, \mnras, 375, L21
\rfr Miglio, A., Montalb\'an, J., \& Dupret, M.-A., 2007b, CoAst, 151, 48
\rfr Morel, T., Butler, K., Aerts, C., et al. \ 2006, \aap, 457, 651 
\rfr Pamyatnykh, A.~A.\ 1999, Acta Astronomica, 49, 119 
\rfr Pamyatnykh, A.~A.\ 2007, CoAst, 150, 207 
\rfr Petersen, J.~O., \& J{\o}rgensen, H.~E.\ 1972, \aap, 17, 367 
\rfr Rogers, F.~J., \& Iglesias, C.~A.\ 1992, \apjs, 79, 507
\rfr Seaton, M.~J., Yan, Y., Mihalas, D., \& Pradhan, A.~K.\ 1994, \mnras, 266, 805 
\rfr Townsend, R.~H.~D., \& MacDonald, J.\ 2006, \mnras, 368, L57 
\rfr Zhevakin, S.~A.\ 1959, Soviet Astronomy, 3, 913 
\rfr Zima, W., Lehmann, H., St{\"u}tz, C.,et al.\ 2007, \aap, 471, 237 

}

\end{document}